\documentclass[twocolumn]{jpsj2} 
\usepackage{bm}

\title{Pressure-Temperature Phase Diagram of Golden SmS}

\author{Keiichiro \textsc{Imura}$^{}$\thanks{E-mail address : imura@mlab2.phys.nagoya-u.ac.jp }, 
Kazuyuki \textsc{Matsubayashi}$^{1}$,
Hiroyuki S. \textsc{Suzuki}$^{2}$,
Noriyuki \textsc{Kabeya}$^{}$,
Kazuhiko \textsc{Deguchi}$^{}$,
 and Noriaki K. \textsc{Sato}$^{}$}

\inst{$^{}$Department of Physics, Nagoya University, Nagoya 464-8602 \\
$^{1}$Institute for Solid State Physics, The University of Tokyo, Kashiwa 277-8581 \\
$^{2}$Nanomaterial Laboratory, National Institute for Materials Science, Tsukuba 305-0047}

\abst{We measured the thermal expansion of the valence fluctuating phase of SmS (golden SmS) to construct a pressure vs temperature phase diagram. The obtained phase diagram is characterized by three lines. One is a crossover line that divides the paramagnetic phase into two regions. The other two lines correspond to a second-order N\'eel transition and a first-order N\'eel transition. The crossover line appears to emerge from a tricritical point that separates the first-order N\'eel transition from the second-order one. We argue that a valence jump occurs at the border of antiferromagnetism.
}

\kword{SmS, valence fluctuation, thermal expansion, high-pressure experiment, phase diagram}

\begin{document}
\maketitle

\section{Introduction}
The hybridization between 4$f$ and conduction electrons forms a heavy-fermion state accompanied by spin and/or charge fluctuations. Although heavy-fermion compounds are basically metallic, they show various ground states, for example, heavy Fermi liquid, magnetic orderings and superconductivity. Interestingly, there are some compounds forming an energy gap or a pseudogap at low temperatures. Typical examples include SmB$_6$ and high-pressure-phase SmS, which we call golden SmS hereafter \cite{101}.
Quantum phase transition (QPT), which is a phase transition at zero temperature, has also attracted our interest, because some exotic phenomena such as non-Fermi liquid and unconventional superconductivity appear in the vicinity of QPT. In heavy fermions, relevant energy scales are so small that QPT is easily tuned by rather small external parameters such as pressure and magnetic field.
In this study, we focus on the pressure-induced phase transition in golden SmS.

At ambient pressure, SmS is a nonmagnetic, ionic crystal with a small energy gap ($E_g \sim$ 0.1 eV). The total angular momentum arising from the Sm$^{2+}$ (4$f^6$) configuration is vanishing ($J = |L-S| = 0$), where $L=3$ and $S=3$ denote the orbital and spin angular momenta, respectively. With increasing pressure, energy gap decreases monotonically and finally collapses at a certain pressure ($P_{c1} \sim$ 7 kbar at room temperature) \cite{Jayaraman, Wachter}. This isostructural first-order phase transition involves a valence change from divalence to mixed valence of Sm$^{2+}$ and Sm$^{3+}$ accompanied by a color change from black to golden yellow. This is the reason why the high-pressure phase is called golden SmS. Note that a pseudogap appears to open in this intriguing phase \cite{Matsubayashi2}. When external pressure is further increased at low temperatures, SmS undergoes a second phase transition at a critical pressure ($P_{c2} \sim$ 19 kbar at $T \sim$ 0).

Improvements in high-pressure experimental techniques have contributed to the elucidation of the electronic state of golden SmS: Experiments on ac specific heat and thermal expansion revealed a sharp anomaly in their temperature dependence, indicating the presence of a bulk phase transition at pressures above $P_{c2}$ \cite{Haga, Imura1}. Nuclear forward scattering (NFS) experiments clearly showed that the phase transition is of magnetic origin\cite{Barla}. This means that the magnetic Sm$^{3+}$ ionic state with the configuration 4$f^5$ is stable at high pressures above $P_{c2}$. Furthermore, ac magnetic susceptibility experiments indicated that the magnetic ground state is antiferromagnetic \cite{Matsubayashi1}.

The NFS experiment showed the sudden appearance of an internal magnetic field when pressure increases across $P_{c2}$, suggesting the first-order phase transition at low temperatures. On the other hand, specific heat and thermal expansion experiments demonstrated the second-order nature of the phase transition at pressures greater than $P_{c2}$. The combination of these two results suggests that there is a tricritical point (TCP) that separates the first-order N\'eel transition from the second-order one near $P_{c2}$ in the pressure ($P$) vs temperature ($T$) phase diagram. However, a complete $P-T$ phase diagram is still missing.
In this paper, therefore, we present a $P-T$ phase diagram of golden SmS using the thermal expansion technique, in which the presence of the TCP is confirmed by the observation of a discontinuous volume change in the vicinity of $P_{c2}$.

From an experimental point of view, thermal expansion is a very useful tool for studying QPT, because it will be largely enhanced near QPT owing to a diverging Gr\"uneisen parameter\cite{3,4}. On the other hand, note that a pressure-transmitting medium has a crucial effect on physical properties including thermal expansion. In our previous report, pressure hysteresis was observed in the $P-T$ phase diagram of SmS above 10 kbar\cite{Imura1}. This pressure roughly coincides with the ``solidification" pressure of Fluorinert that was used as a pressure-transmitting medium, and hence the hysteresis may not be intrinsic. In this study, we first determined the solidification pressure of Daphne oil 7373 as a function of temperature, and then remeasured the thermal expansion of SmS single crystals using Daphne oil to obtain the intrinsic effects of pressure on SmS.

\section{Experimental Results and Discussion}

\subsection{Experimental methods}
SmS single crystals were grown from starting materials of 99.99$\%$ pure (4N) Sm chips and 6N powdered S by the vertical Bridgman method in a high-frequency induction furnace \cite{Matsubayashi3}. Three types of single crystals with different starting compositions (nominally stoichiometric, 1$\%$ excess of Sm, and 1$\%$ excess of S samples) were prepared. A sample with typical dimensions of 1.5 $\times$ 1.6 $\times$ 0.45 mm$^3$ was cleaved for measurements. Thermal expansion results were found to be only weakly sample-dependent, so that we report here the results of nominally stoichiometric samples.

Thermal expansion measurement under high pressure was carried out by the active-dummy method using strain gauges (SKF-5414, KYOWA). A copper block (6N purity, typical dimensions of 1.5 $\times$ 1.5 $\times$ 1.0 mm$^3$) was used as a dummy sample. We employed an ac method in terms of a lock-in amplifier to reduce the drift voltage arising from thermoelectric power. An ac voltage with an amplitude of 0.05 V and a frequency of 33 Hz was applied to the strain gauges. Compared with our previous data taken by a dc method, the signal-to-noise ratio was markedly improved \cite{Imura3}.

When external pressure was applied, the pressure cell (a NiCrAl - BeCu hybrid piston cylinder cell) was heated to about 320 K to reduce inhomogeneity arising from the solidification of Daphne oil (see discussion below for details). The pressure at low temperatures was determined by the superconducting transition temperature of indium.

\subsection{$P - T$ phase diagram of Daphne oil 7373}
Inhomogeneity will exert a harmful effect upon SmS. Since the pressure inhomogeneity is induced by the solidification of the pressure-transmitting medium, it is important to know the ``frozen-melted'' phase diagram of Daphne oil 7373. We start with the construction of its $P - T$ phase diagram using SmS as a test material.

\begin{figure}
\begin{center}
\includegraphics[width=90mm]{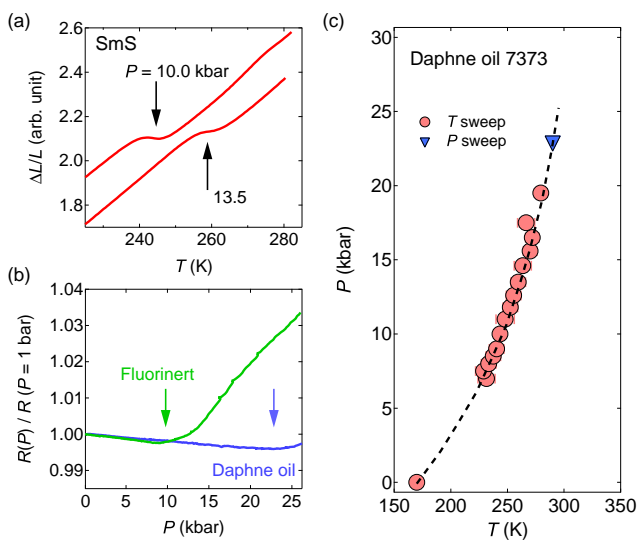}
\end{center}
\caption{ (Color online) (a) Temperature dependence of thermal expansion $\Delta L$/$L$($T$) of SmS at 10.0 and 13.5 kbar measured during heating. The anomalies denoted by arrows are due to the solidification of Daphne oil 7373. (b) Pressure dependence of the relative electrical resistivity of a strain gauge at room temperature. Here, the gauge was attached to the reference sample Cu, and Fluorinerts FC70 and FC77 and Daphne oil 7373 were used as pressure-transmitting media. The arrows show anomalies due to the solidification of the pressure-transmitting media. (c) Freezing temperature of Daphne oil 7373. The circles and triangles were deduced from the temperature and pressure dependences of thermal expansion and electrical resistivity shown in (a) and (b), respectively. The broken line is a visual guide.}
\label{D}
\end{figure}

Figure \ref{D}(a) shows the temperature dependence of the thermal expansion $\Delta L/L(T)$ of SmS at 10 and 13.5 kbar. The anomaly observed at approximately 250 K is due to the freezing of Daphne oil. This enables us to determine the ``solidification pressure'' or equivalently the ``melting temperature'' of Daphne oil.

The solidification pressure is shown in Fig.~\ref{D}(c) as a function of temperature. When Daphne oil is kept at room temperature, it does not solidify even at 20 kbar. 
Reflecting this nature of Daphne oil, the electrical resistivity of a strain gauge, which was attached to the reference sample Cu, shows no anomaly up to about 22 kbar, as shown in Fig.~\ref{D}(b). By contrast, the same measurement using Fluorinert as a pressure-transmitting medium shows an anomaly at approximately 10 kbar, which we ascribe to the solidification of Fluorinert\cite{22}.

The $P - T$ phase diagram of Daphne oil tells us that pressure should be applied at the highest temperature possible. This is the reason why we heated the pressure cell when applying pressure.
This is also effective in the case of Daphne oil 7474\cite{Murata}.

\subsection{$P - T$ phase diagram of golden SmS}

\begin{figure}
\begin{center}
\includegraphics[width=90mm]{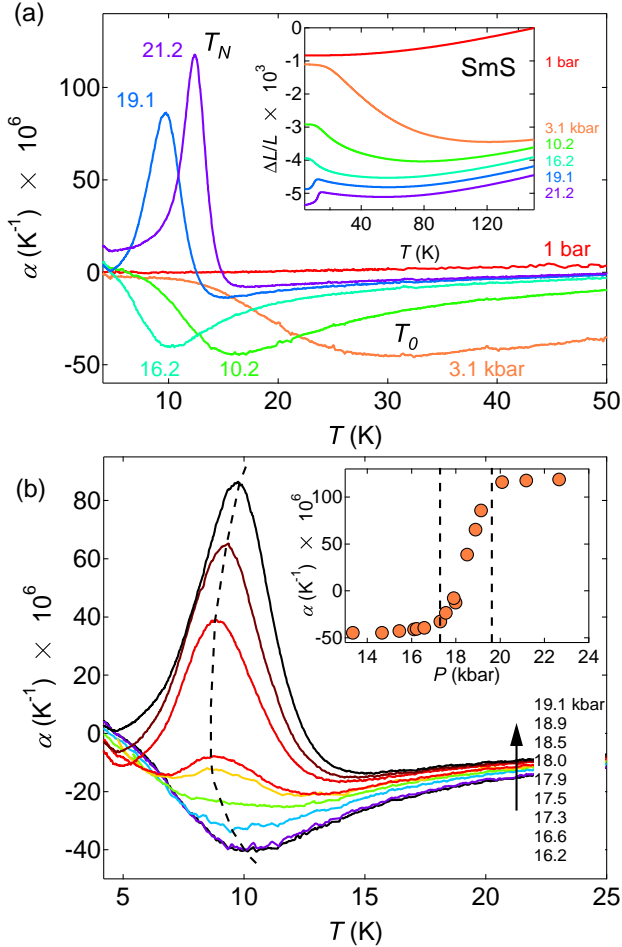}
\end{center}
\caption{ (Color online) (a) Temperature dependences of thermal expansion coefficient $\alpha(T)$ of SmS at selected pressures up to 21.2 kbar. $T_0$ and $T_N$ are defined as the temperatures at which a shallow minimum and a positive, rather sharp peak appear, respectively. The inset shows the temperature dependence of $\Delta L$/$L$ at the same pressure as that of $\alpha(T)$. Data at different pressures are shifted for clarity. (b) Temperature dependences of thermal expansion coefficient at selected pressures between 16.2 and 19.1 kbar. The inset shows $\alpha(T)$ along the broken line in the main frame as a function of pressure. The region sandwiched by the two broken lines denotes a two-phase mixture region.}
\label{L}
\end{figure}

Figure \ref{L}(a) shows the temperature dependence of the thermal expansion coefficient $\alpha(T)$ obtained from the thermal expansion $\Delta L$/$L(T)$ given in the inset via the relation $\alpha(T) \equiv ({\rm d}L/{\rm d}T)/L$. At ambient pressure (black phase), the sample length $L$, proportional to the volume $V$ for the cubic crystal, decreases monotonically with decreasing temperature (see the inset). At 3.1 kbar (golden phase), $\Delta L$/$L(T)$ exhibits a strikingly different behavior from that of the black phase; the volume increases with decreasing temperature below about 130 K. 
Corresponding to this unusual behavior in the $\Delta L$/$L$ curves of the golden phase, $\alpha(T)$ exhibits a broad minimum ($\sim$ $-$4.5 $\times$ 10$^{-5}$) at a temperature $T_0$.
Note that this minimum value is almost independent of pressure.
The peak structure is described by a Schottky model and the characteristic temperature $T_0$ decreases with pressure. These results are consistent with our previous reports\cite{Matsubayashi2, Imura3}, indicating that there is no distinction between the data obtained for Daphne oil and Fluorinert in the pressure range tested ($P <$ 16.2 kbar), although the pressure range exceeds the solidification pressure of Fluorinert ($\sim 10$ kbar).

At pressures above $P_{c2} \sim$ 19 kbar, a kink anomaly appears in the dilatation $\Delta L$/$L(T)$ curve at a low temperature.
Corresponding to that, a sharp anomaly occurs in the $\alpha(T)$ curve, which signals a N\'eel transition, $T_N$. The positive sign of the anomaly implies that $T_N$ increases with pressure, consistent with the present results and previous data of  specific heat and magnetic susceptibility \cite{Haga,Matsubayashi1}. 

Figure~\ref{L}(b) shows the temperature dependences of thermal expansion coefficient at different pressures between 16.2 and 19.1 kbar. Interestingly, we observe a peak with a positive sign; the peak is superposed on the broad minimum situated at about 10 K and grows with increasing pressure along the broken line. It is reasonable to ascribe the observation to a  two-phase mixture consisting of the broad minimum and positive peak arising from the paramagnetic and antiferromagnetic states, respectively.

In the inset of Fig.~\ref{L}(b), we plot $\alpha$ along the broken line. This is interpreted as the volume fraction of the antiferromagnetic phase, which steeply increases at a pressure interval $\Delta P \sim 1.8$ kbar between 17.3 and 19.1 kbar.

\begin{figure}
\begin{center}
\includegraphics[width=70mm]{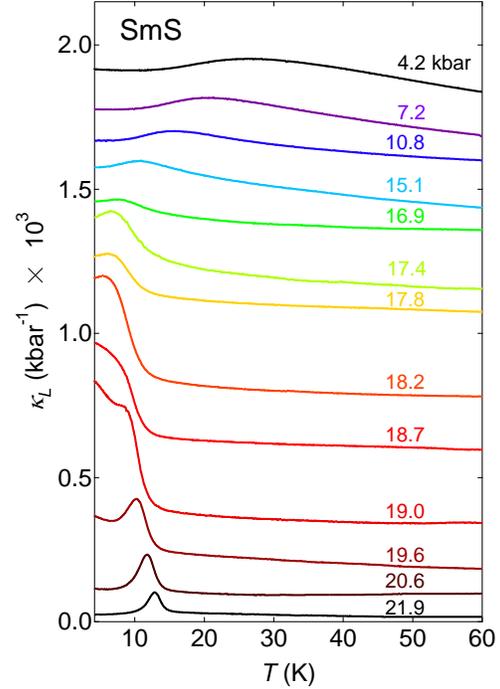}
\end{center}
\caption{ (Color online) Temperature dependences of the compressibility $\kappa_L(T)$ at selected pressures. The broad peak below 17.8 kbar and the sharp peak above 19.6 kbar correspond to $T_0$ and $T_N$ deduced from $\alpha(T)$, respectively. Data at different pressures are shifted for clarity. See text for details.}
\label{C}
\end{figure}

Figure \ref{C} shows the temperature dependences of the compressibility $\kappa_L(T)$ at selected pressures, which were obtained using the relation $\kappa_L = -\{\Delta L/L|_{P_2}-\Delta L/L|_{P_1}\}/(P_2-P_1)$. At low pressures, we only observe a broad peak corresponding to the anomaly of $\alpha(T)$ at $T_0$. As pressure increases, the peak structure becomes prominent. At pressures higher than $P_{c2}$, e.g., 19.6 kbar, a sharp peak appears at the N\'eel temperature.

\begin{figure}
\begin{center}
\includegraphics[width=85mm]{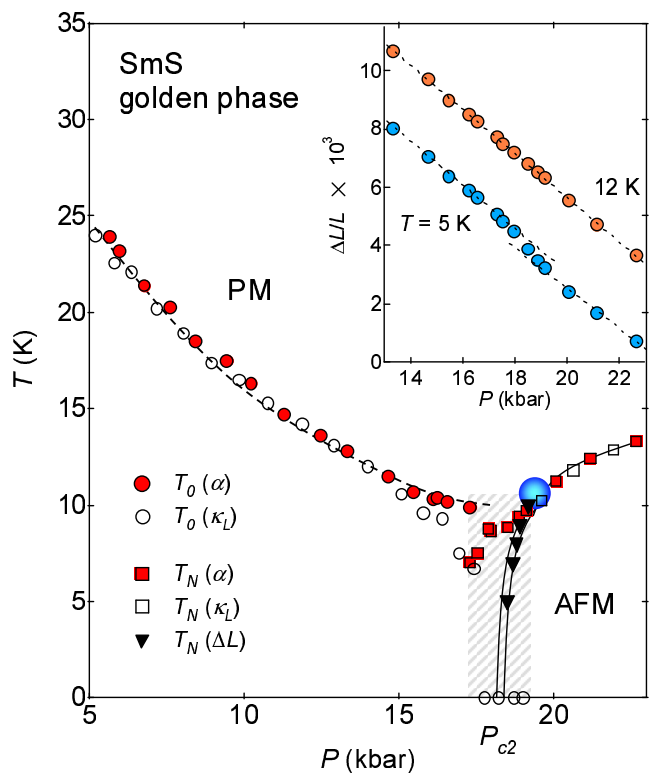}
\end{center}
\caption{(Color online) $P - T$ phase diagram of golden SmS. Circles and squares were deduced from $\alpha(T)$ and $\kappa_L(T)$ shown in Figs. 2 and 3, respectively. The broken line denotes a crossover at $T_0$, and the solid and double solid lines correspond to the second-order and first-order N\'eel transitions, respectively. The hatched region at approximately $P_{c2}$ shows the two-phase mixture region that is possibly caused by sample inhomogeneity.
The inset shows the pressure dependences of the thermal expansion $\Delta L$/$L$($P$) at 5 and 12 K. Note that the $\Delta L$/$L$($P$) curve shows a jump at approximately 18.5 kbar. This anomaly defines the first-order N\'eel transition plotted in the phase diagram. The large circle shows the tricritical point ($P_{t}$, $T_{t}$).}
\label{PT}
\end{figure}

From these anomalies, we construct the $P - T$ phase diagram of golden SmS, as shown in Fig.~\ref{PT}. The broken line through the open and closed circles denotes a crossover dividing the paramagnetic phase into the low- and high-temperature regions. The solid line through the open and closed squares indicates the second-order N\'eel transition, and the double solid line through the triangles corresponds to the first-order N\'eel transition, as shown below. The large circle denotes the TCP that separates the first-order transition from the second-order transition. The hatched region indicates the two-phase mixture region mentioned above. Interestingly, note that the crossover line $T_0(P)$ (deduced from $\alpha(T)$) appears to terminate at the TCP, although the data points near TCP deduced from $\kappa_L$ deviate from it presumably owing to pressure inhomogeneity. According to Matsubayashi {\it et al.}, $T_0$ is the characteristic temperature of the formation of a pseudogap \cite{Matsubayashi2}.

To estimate dilatation as a function of pressure $\Delta L$/$L(P)$, instead of temperature, we need to know the absolute value of the compressibility $\kappa$ (or equivalently the bulk modulus $B$) at a certain temperature lower than the melting temperature of Daphne oil, e.g., at 150 K (see Fig.~1c). However, we only have available data taken at room temperature. Therefore, we assume that $B$($T$ = 150 K) = $B$($T$ = 300 K) \cite{9}. Any other choice of $B$ does not change the following conclusion. The thus-evaluated $\Delta L$/$L(P)$ at a low temperature is shown in the inset of Fig.~\ref{PT}.
At 5 K, $\Delta L$/$L$($P$) shows a steplike, albeit obscured by some inhomogeneity (see below), anomaly at approximately 18 kbar. By contrast, no jump is visible at 12 K; we only observe the $P$-linear behavior of $\Delta L$/$L$($P$). 
This means that the TCP lies between 5 and 12 K. From a detailed study of the $T$ dependence of the jump $\Delta V  $\cite{ImuraD}, we determined the position of the TCP on the $P - T$ plane to be ($P_t$, $T_t$) = (19.0 $\pm$ 0.8 kbar, 10.5 $\pm$ 0.5 K), as shown in Fig.~\ref{PT}. Note that the compressibility exhibits a very large anomaly near the TCP (Fig.~\ref{C}).
We also determine $P_{c2}$ by the extrapolation of the first-order line to zero temperature as $P_{c2}$ = 18.1 $\pm$ 0.8 kbar.

Strictly speaking, the volume change observed is not a jump but a rapid variation at a narrow pressure interval. Note that this transient region between 17.5 and 19.1 kbar almost coincides with the $\Delta P$ mentioned above (see the inset of Fig.~\ref{L}(b)). Therefore, the broadened nature of the volume change is ascribed to the two-phase mixture arising from the inhomogeneity of pressure and/or the sample.

According to our thermal expansion study of CeRhIn$_5$ and UGe$_2$, the pressure gradient of Daphne oil 7373 is estimated at less than 0.2 kbar \cite{chen,10}. This is much smaller than the pressure interval $\Delta P$ mentioned above. We consider that the wide transient region is attributed to sample inhomogeneity rather than to pressure inhomogeneity. As mentioned in Introduction, the pressure-induced black-to-golden phase transition at $P_{c1}$ is accompanied by a large volume change that amounts to about 10\%. Therefore, it is likely that some inhomogeneity is induced in the sample during the black-to-golden phase transition.

\subsection{Pressure dependence of residual resistivity}

\begin{figure}
\begin{center}
\includegraphics[width=90mm]{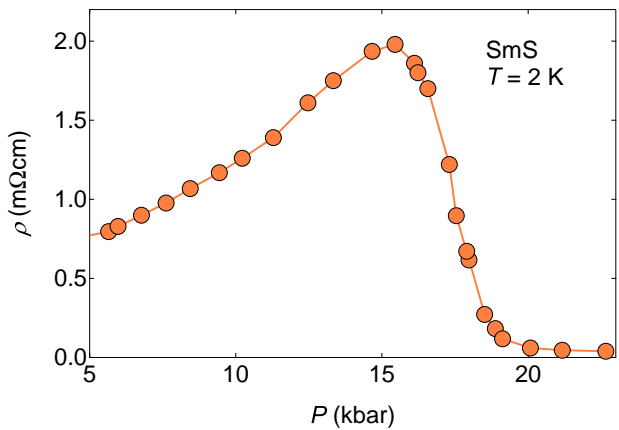}
\end{center}
\caption{ (Color online) Pressure dependence of electrical resistivity at 2 K. It shows a maximum at approximately 15.5 kbar, which does not coincide with $P_{c2}$ (= 18.1 $\pm$ 0.8 kbar). Note that the data points above $P_{c2}$ are correspond to the antiferromagnetic phase.}
\label{R}
\end{figure}

Let us discuss in more detail the volume jump in conjunction with valence change. In general, a Sm$^{2+}$ ion has a greater ionic radius than a Sm$^{3+}$ ion. Furthermore, a conduction electron, which can be produced as a result of the promotion of a 4f electron from a Sm$^{2+}$ ion into a conduction band, tends to expand its wave function so as to obtain a kinetic energy gain, resulting in a reduction in crystal volume. As a result, it is expected that a semiconducting phase composed of Sm$^{2+}$ ions has a larger volume than a metallic phase composd of Sm$^{3+}$ ions. 
In the present case, therefore, it is very likely that the average volume is larger in the paramagnetic phase with the mixed valence than in the antiferromagnetic phase with the metallic conductivity that mainly consists of Sm$^{3+}$ ions. This suggests that valence change occurs at the border of antiferromagnetism. 

According to theory, a valence change gives rise to an enhancement of residual resistivity\cite{Miyake}. This is because impurity potential is enhanced by a many-body effect in the vicinity of a quantum critical point of valence transition at which the correlation length of valence fluctuations diverges. Therefore, we expect that the residual resistivity of golden SmS shows a maximum at approximately $P_{c2}$.

To check this, we measured electrical resistivity under pressure. The detailed results will be published elsewhere. In Fig.~\ref{R}, we show only the pressure dependence of the electrical resistivity taken at 2 K, which is tentatively regarded as the residual resistivity $\rho_0$. It is clearly seen that $\rho_0$ initially increases with pressure and then passes through a maximum before steeply decreasing, consistent with previous reports \cite{Lapierre, Koncykowski}. It is possible to speculate that the initial increase in $\rho_0$ is ascribed to the decrease in mobility from the increase in carrier concentration at 300 K with pressure\cite{Mizuno}.

As is expected, $\rho_0$ shows a maximum near $P_{c2}$. However, we note that the maximum pressure does not exactly coincide with $P_{c2}$. There are two possible explanations for this deviation. First, the phase transition at $P_{c2}$ is of the first-order; therefore, the theoretical model mentioned above may not be applicable to the present case. Second, the existence of the transient region gives rise to not only the broadening of the peak structure but also the shift in peak position. It is unclear at present which is more probable. Thus, we need further investigation.

 \section{Conclusions}

We first constructed a phase diagram of Daphne oil 7373. The freezing temperature at 25 kbar was estimated as about 300 K. Therefore, we pressurized the cell at 320 K to avoid the inhomogeneity induced by the solidification of the pressure-transmitting medium.

We measured the thermal expansion of golden SmS at pressures of up to 22.7 kbar, and observed a broad peak at the characteristic temperature $T_0$ and a sharp peak at the N\'eel temperature $T_N$. From these anomalies, we constructed the $P$-$T$ phase diagram. 
It consists of three characteristic lines: the crossover line corresponding to $T_0$, and the second-order and first-order phase lines corresponding to $T_N$. These lines appear to emerge from the tricritical point, located at ($T_t$, $P_t$) $\sim$ (19.0 $\pm$ 0.8 kbar, 10.5 $\pm$ 0.5 K) on the pressure vs temperature plane. A linear extrapolation of the first-order phase line provides the critical pressure $P_{c2} = 18.1 \pm 0.8 $ kbar.
In the vicinity of $P_{c2}$, there is a two-phase-mixture region, in which both the paramagnetic and antiferromagnetic phases are mixed owing to sample inhomogeneity.

We observed a volume jump at $P_{c2}$ where the quantum phase transition between the paramagnetic and antiferromagnetic phases occurs. We measured residual resistivity as a function of pressure, and found that it shows a maximum at a pressure near $P_{c2}$. Considering the correlation of volume with valence, we pointed out the possible valence jump at the border of antiferromagnetism.

\section*{Acknowledgments}
We thank S. Kimura, T. Ito, T. Saso, T. Kasuya, and J. Flouquet for helpful comments.
This work was partially supported by KAKENHI (S) (No. 20224015). One of the authors (K. I.) was supported by a Grant-in-Aid for JSPS Fellows.


\begin{thebibliography}{99} 

\bibitem{101} N. F. Mott: Philos. Mag. {\bf 30} (1974) 403.

\bibitem{Jayaraman} A. Jayaraman, V. Narayanamurti, E. Bucher, and R. G. Maines: Phys. Rev. Lett. {\bf 25} (1970) 1430.

\bibitem{Wachter} P. Wachter: in {\it Handbook on the Physics and Chemistry of Rare Earths}, ed. K. A. Gschneidner, Jr., L. Eyring, G. H. Lander, and G. R. Choppin (North-Holland, Amsterdam, 1994) Vol. 19, p. 383.

\bibitem{Matsubayashi2} K. Matsubayashi, K. Imura, H. S. Suzuki, G. F. Chen, N. M\^{o}ri, T. Nishioka, K. Deguchi, and N. K. Sato: J. Phys. Soc. Jpn. {\bf 76} (2007) 033602.

\bibitem{Haga} Y. Haga, J. Derr, A. Barla, B. Salce, G. Lapertot, I. Sheikin, K. Matsubayashi, N. K. Sato, and J. Flouquet: Phys. Rev. B {\bf 70} (2004) 220406.

\bibitem{Imura1} K. Imura, K. Matsubayashi, H. S. Suzuki, T. Nishioka, N. M\^{o}ri, and N. K. Sato: Physica B {\bf 378-380} (2006) 728.

\bibitem{Barla} A. Barla, J. P. Sanchez, Y. Haga, G. Lapertot, B. P. Doyle, O. Leupold, R. R\"{u}ffer, M. M. Abd-Elmeguid, R. Lengsdorf, and J. Flouquet: Phys. Rev. Lett. {\bf 92} (2004) 066401.

\bibitem{Matsubayashi1} K. Matsubayashi, K. Imura, H. S. Suzuki, S. Ban, G. F. Chen, K. Deguchi, and N. K. Sato: J. Magn. Magn. Mater. {\bf 310} (2007) 408.

\bibitem{3} L. Zhu, M. Garst, A. Rosch, and Q. Si: Phys. Rev. Lett. {\bf 91} (2003) 066404.

\bibitem{4} R. K\"{u}chler, N. Oeschler, P. Gegenwart, T. Cichorek, K. Neumaier, O. Tegus, C. Geibel, J. A. Mydosh, F. Steglich, L. Zhu, and Q. Si: Phys. Rev. Lett. {\bf 91} (2003) 066405.

\bibitem{Matsubayashi3} K. Matsubayashi, K. Imura, H. S. Suzuki, T. Mizuno, S. Kimura, T. Nishioka, K. Kodama, and N. K. Sato: J. Phys. Soc. Jpn. {\bf 76} (2007) 064601.

\bibitem{Imura3} K. Imura, K. Matsubayashi, H. S. Suzuki, K. Deguchi, and N. K. Sato: to be published in Physica B.

\bibitem{22} T. Sakai, T. Kagayama, and G. Oomi: J. Mater. Process. Technol. {\bf 85} (1999) 224.

\bibitem{Murata} K. Murata, K. Yokogawa, H. Yoshino, S. Klotz, P. Munsch, A. Irizawa, M. Nishiyama, K. Iizuka, T. Nanba, T. Okada, Y. Shiraga, and S. Aoyama: Rev. Sci. Instrum. {\bf 79} (2008) 085101.

\bibitem{9} R. Keller, G. G\"{u}ntherodt, W. B. Holzapfel, M. Dietrich, and F. Holtzberg: Solid State Commun. {\bf 29} (1979) 753.

\bibitem{ImuraD} K. Imura: Dr. Thesis, Graduate School of Science, Nagoya University, Nagoya (2008).


\bibitem{chen} G. F. Chen, K. Matsubayashi, S. Ban, K. Deguchi, and N. K. Sato: Phys. Rev. Lett. {\bf 97} (2006) 017005.

\bibitem{10} N. Kabeya, R. Iijima, E. Osaki, S. Ban, K. Imura, K. Deguchi, N. Aso, Y. Honma, Y. Shiokawa, and N. K. Sato: to be published in Physica B.

\bibitem{Miyake} K. Miyake, O. Narikiyo, and Y. Onishi: Physica B {\bf 259-261} (1999) 676-677.

\bibitem{Lapierre} F. Lapierre, M. Ribault, F. Holtzberg, and J. Flouquet: Solid State Commun. {\bf 40} (1981) 347.

\bibitem{Koncykowski} K. Konczykowski, J. Morillo, and J. P. Senateur: Solid State Commun. {\bf 40} (1981) 517.

\bibitem{Mizuno} T. Mizuno, T. Iizuka, S. Kimura, K. Matsubayashi, K. Imura, H. S. Suzuki, and N. K. Sato: J. Phys. Soc. Jpn. {\bf 77} (2008) 113704.


\end{thebibliography}
\end{document}